# Atomistic study of hardening mechanism in *Al-Cu* nanostructure

Satyajit Mojumder [a,*], Tawfiqur Rakib [a], Mohammad Motalab[a], Dibakar Datta[b]

[a]Department of Mechanical Engineering, Bangladesh University of Engineering and Technology, Dhaka-1000, Bangladesh.

[b] Department of Mechanical and Industrial Engineering, New Jersey Institute of Technology (NJIT), Newark, NJ 07102, USA.

**Abstract:**

Nanostructures have the immense potential to supplant the traditional metallic structure as they show enhanced mechanical properties through strain hardening. In this paper, the effect of grain size on the hardening mechanism of *Al-Cu* nanostructure is elucidated by molecular dynamics simulation. *Al-Cu* (50-54% *Cu* by weight) nanostructure having an average grain size of 4.57 to 7.26 nm are investigated for tensile simulation at different strain rate using embedded atom method (EAM) potential at a temperature of 50~500K. It is found that the failure mechanism of the nanostructure is governed by the temperature, grain size as well as strain rate effect. At the high temperature of 300-500K, the failure strength of *Al-Cu* nanostructure increases with the decrease of average grain size following Hall-Petch relation. Dislocation motions are hindered significantly when the grain size is decreased which play a vital role on the hardening of the nanostructure. The failure is always found to initiate at a particular *Al* grain due to its weak link and propagates through grain boundary (GB) sliding, diffusion, dislocation nucleation and propagation. We also visualize the dislocation density at different grain size to show how the dislocation affects the material properties at the nanoscale. These results will further aid investigation on the deformation mechanism of nanostructure.

---

*Corresponding author: Tel: +880-1737-434034, E-mail address: satyajit@me.buet.ac.bd

**Keywords:** *Al-Cu* nanostructure, Grain boundary, hardening, Dislocations.

## 1. Introduction

The exceptional mechanical behavior of nanostructure materials showing enhanced ductility[1,2], increased strength and hardness[3] attracted the interest of many researchers in recent days. These interesting behaviors arise from the intricate interplay between dislocation and grain size. In this context, grain size has a dominant role over the mechanical properties of nanostructure[4,5]. So, these nanostructures are also nanocrystalline in nature. The plastic deformation mode of coarse-grained materials is suppressed in nanocrystalline materials arising a complex deformation process. These include grain boundary(GB) evolution, dislocation phenomenon, and nanocrystal plasticity[6].

Due to the improved characteristics of nanostructures, Aluminum (*Al*) and Copper (*Cu*), two widely used engineering materials have been already extensively used in automobile, packaging and construction industries. Recently, due to their immense potential in the different application of the nanoscale device, *Al* and *Cu* are investigated by the researchers in the form of nanocrystalline structure [7,8]. It is well established that the nanostructures of *Al* and *Cu* in their pure or alloy form have the superior mechanical strength and elastic property [9,10], electrical property [11,12], and thermal property[13,14]. Also, the inclusion of *Al* and *Cu* in the same structure is found to show improved corrosion resistance and high strength [15]. However, to the best of the knowledge of the authors, no attempt has been taken by the researchers up to now to amalgamate these two nanostructures (*Al* and *Cu* nanocrystalline grain) as a combined structure which should have unique mechanical properties compared to the individual nanocrystalline structure of *Al* and *Cu*.

The failure of nanostructures is always very intriguing as the mechanism is driven by the intertwine of GB sliding, rotation, diffusion, dislocation nucleation, dislocation elastic field, etc. Besides, the grain size also affects the failure mechanism and hardening of material according to the Hall-Petch phenomenon. The existence of the Hall-Petch effect is found in many engineering materials [16,17] which indicate that with smaller grain size, the materials show higher strength. Hansen [16] showed that the dislocation based mechanisms are the main key factors at all scale to demonstrate the Hall-Petch effect. However, when the grain size is reduced to nano-level, this effect is inversed and this phenomenon is termed as inverse Hall-Petch effect. Conrad and Narayan [18] observed softening or inverse Hall-Petch effect for the polycrystalline materials of grain size less than a critical value of 10-50 nm. The explanation behind this reverse behavior can be attributed to dislocation based models, diffusion based models, and grain boundary (GB) shearing based models. Carlton et al. [19] suggested a model which assumes the probability of the atoms in dislocation core to be absorbed by the grain boundaries and gives rise to this reverse hardening mechanisms. Recent studies also reported the presence of this effect in nanocrystalline metals[20,21]. However, for Al-Cu nanostructure, the mechanism is not that straightforward and not well-known. Hence a thorough investigation is necessary.

On many applications like welding and casting [22], metals undergo different temperatures and loading rates, which severely affects their strength and mechanical properties [23]. Lee et al. [24] investigated the effect of temperature and strain rate experimentally on the dynamic impact properties of 7075 Aluminum alloy and concluded that the compressive stress-strain response of the material is affected by temperature and strain rate. Elia et al. [25] reported that cooling rate has a tremendous impact on the grain refinement of B-206 *Al-Cu* alloys. Therefore, temperature and strain rate dependencies should be studied to depict the mechanical properties of *Al-Cu* nanostructures too.

Keeping the above-mentioned facts in mind, the primary objective of this paper is to model the *Al-Cu* nanostructure and investigate the grain size and temperature effects on its mechanical properties at the nanoscale by molecular dynamics simulations. The impact of strain rate variation on the mechanical properties is also studied. Finally, the influence of grain boundaries on the dislocation activities and the underlying deformation mechanisms of *Al-Cu* nanostructure under uniaxial tensile loading are elucidated.

## 2. Methodology

The uniaxial tensile simulations are performed for *Al-Cu* nanostructure having a dimension of 10nm×10 nm×20 nm in X, Y and Z direction respectively using EAM [26] potential. This EAM potential is widely used to define atomic interactions for aluminum, copper and other metals in its nanocrystalline forms [27–29]. The model contains *Al* and *Cu* grain where the grain numbers are controlled but randomly oriented. The grain number ratio of *Al* and *Cu* are maintained as 7:3 which makes a percentage of *Cu* as 50-54% by weight. The grain number is varied from 10 to 40 with a grain size (diameter) of 4.57 to 7.26 nm, and the total number of atoms is around 140000. The *Al-Cu* nanostructure with *Al* and *Cu* grain is shown in Fig. 1. The grain-based *Al-Cu* nanostructure is generated using APoGe Software package[30]. The initial geometry is relaxed sufficiently (for 100 ps) under the NPT dynamics. Later, a tensile load is applied at different temperatures (50K, 100K, 300K and 500K) along the Z direction (see Fig. 1) of the simulation box at a strain rate of $10^9$ s$^{-1}$. The timestep chosen for all the simulation is 1fs. The strain rate is also varied to investigate its effect on the failure behavior. The tensile simulations are performed using LAMMPS Package[31], and post-processing is done using OVITO[32] software package.

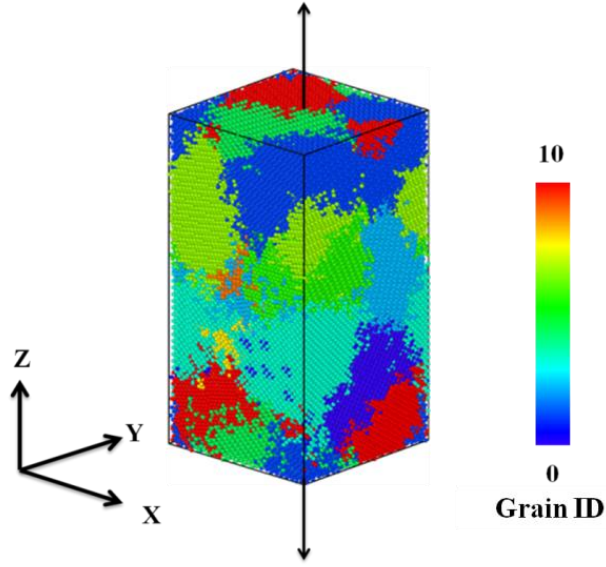

Fig. 1: Al-Cu nanostructure with 7.2 nm grain size. The coordinate system is shown in the figure and uniaxial tensile loading is applied along the Z direction shown with an arrowhead.

For obtaining the stress-strain behavior, atomic stresses are calculated as the simulation box is deformed uniaxially. Atomic stresses were calculated based on the definition of virial stress, which is expressed as[33,34]

$$\sigma_{virial}(r) = \frac{1}{\Omega}\Sigma_i \left[ \left(-m_i \dot{u}_i \otimes \dot{u}_i + \frac{1}{2}\Sigma_{j \neq i} r_{ij} \otimes f_{ij}\right)\right] \quad (1)$$

where, the summation is over all the atoms occupying the total volume $\Omega$, $\otimes$ indicates the cross product, $m_i$ is the mass of atom $i$, $r_{ij}$ is the position vector of atom, $\dot{u}_i$ is the time derivative which indicates the displacement of an atom with respect to a reference position, and $f_{ij}$ is the interatomic force applied on atom *i* by atom *j*.

To investigate the stability of *Al-Cu* nanostructures, we calculated the optimized energy of the system and compared with the optimized energy of the aluminum and copper nanocrystalline systems. For this, we modeled nanocrystalline *Al* and *Cu* individually having grain size similar to the grain size of *Al-Cu* nanostructure used in the present study. The results of the optimized energy of these three (nanocrystalline *Al*, *Cu*, and *Al-Cu* nanostructure) are presented in Table 1. The optimized energy of *Al-Cu* nanostructure lies between the optimized energy for nanocrystalline *Al* and *Cu* (previously fabricated experimentally[35,36]). Hence the structure of *Al-Cu* nanostructures are energetically feasible to form.

**Table 1:** The optimized energy of nanocrystalline *Al*, *Cu* and *Al-Cu* nanostructure for different grain size.

| Types of System | Optimized energy for different grain size (eV/atom) | | | |
|---|---|---|---|---|
| | 4.57 nm | 5.03 nm | 5.76 nm | 7.26 nm |
| Nanocrystalline copper | -3.47 | -3.47 | -3.49 | -3.48 |
| Nanocrystalline Aluminum | -3.27 | -3.28 | -3.29 | -3.29 |
| Al-Cu nanostructure | -3.39 | -3.41 | -3.4 | -3.39 |

## 3. Results and Discussion

### 3.1 Effect of temperature and grain size

The simulated stress-strain curve for different temperature and average grain size is shown in Fig. 2. At the early stage of loading, the stress and strain relation is linear and the material exhibits elastic behavior. This linear behavior is observed up to a strain value of 0.02 and after that the stress-strain behavior becomes nonlinear. This non-linear material behavior continues to develop for the material until it reaches a peak value which is called the ultimate stress of the nanostructure. After reaching the ultimate stress, there is a sudden drop in the stress due to the formation of crack (around 7% strain) inside the material. Due to the ductile nature of the metallic nanostructure, it does not fail catastrophically during the sudden load drop. Afterwards, the flow stress is visualized which is due to plasticity in the metallic nanostructure material. Another interesting finding is that the *Al-Cu* nanostructure shows higher strength than nanocrystalline aluminum and copper structures. While *Al-Cu* nanostructure possesses around 4 GPa strength, the strength of *Al* and *Cu* are respectively around 0.8 GPa [27] and 2 GPa [37]. This is because the single crystal Al and Cu grains in the nanostructure forms an intermetallic compound. Intermetallic compounds are generally stronger than their individual counterparts. The effects of average grain size and temperature are prominent in the stress-strain diagram and they also govern the mechanical properties of the nanostructure. From Figs. 2(a) and 2(b), it can be seen that the ultimate stress is the highest for the grain size of 5.03 nm at 50K and 100 K temperature. However, Fig 2(c) and 2(d) depicts that the ultimate stress is the highest for lower grain size of 4.57 nm at a temperature of 300K and 500K. At 300K and 500K the ultimate stress shows a decreasing trend with the increase of grain size. On the other hand, a notable finding is that the Young moduli of the material do not vary significantly with the temperature and grain size for *Al-Cu* nanostructures.

It is also observed from the Fig. 2 that the ultimate stress of the nanostructure decreases with the increase of temperature irrespective of different grain size. The nanostructure shows the ultimate stress of around 5GPa at 50K and it reduces to 3.5 GPa at 500K temperature. A similar trend is previously found in an experimental investigation of Aluminum alloy[38]. At high temperature, the nanostructure is observed to behave in more ductile nature. As a result, fracture strain is found to increase while the ultimate stress decreases at high temperature.

The variation of the ultimate stress of *Al-Cu* nanostructure with the grain size is shown in Fig. 3. At the high temperature of 300K and 500K, the ultimate stress decreases with the grain size. Therefore, the nanostructure manifests its Hall-Petch effect. As the grain size becomes small, the nanostructure experiences the hardening phenomenon. Metal nanocrystals generally show an inverse Hall-Petch effect when the grain size is lower than 10 nm[20].

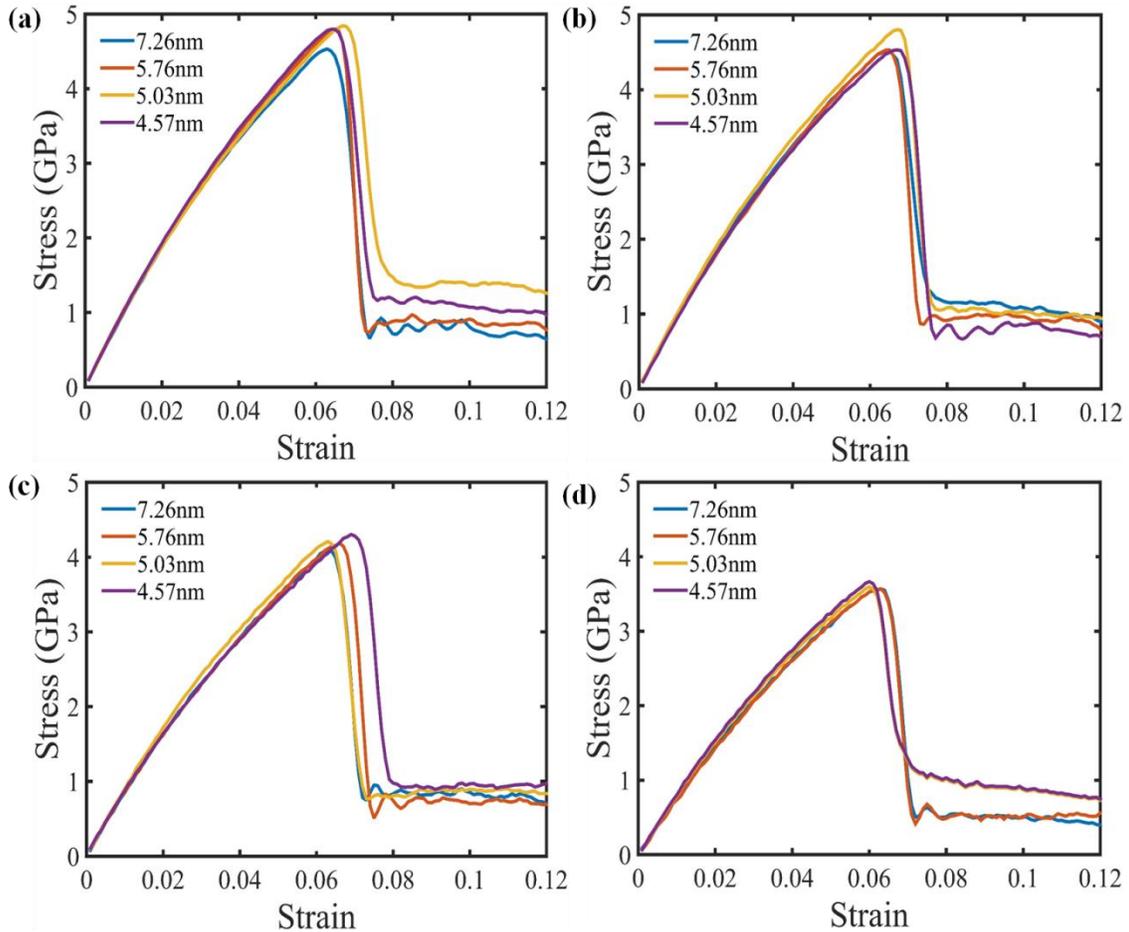

Fig. 2: Stress-strain diagram for the tensile test simulation of *Al-Cu* nanostructure for different average grain size at (a) 50K, (b) 100K, (c) 300K and (d) 500K.

On the contrary, in the case of *Al-Cu* nanostructure nanowire, we have observed that Hall-Petch relationship exists at a high temperature of 300-500K. We fit the obtained data with the general Hall-Petch relations-

$$\sigma_y = \sigma_0 + \frac{K}{\sqrt{d}} \tag{2}$$

Where $\sigma_y$ is the yield stress, $\sigma_o$ is a material constant for the starting stress for dislocation movement (or the resistance of the lattice to dislocation motion), $k$ is the strengthening coefficient (a constant specific to each material), and $d$ is the average grain size. The value of $\sigma_o$ is 3.59, 3.236 GPa and $K$ is 1.492, 0.8398 at 300K and 500K, respectively. For low temperature, inverse Hall-Petch relation is found after a critical grain size of 5.03 nm. The details of fracture mechanisms are discussed in Section 3.2.

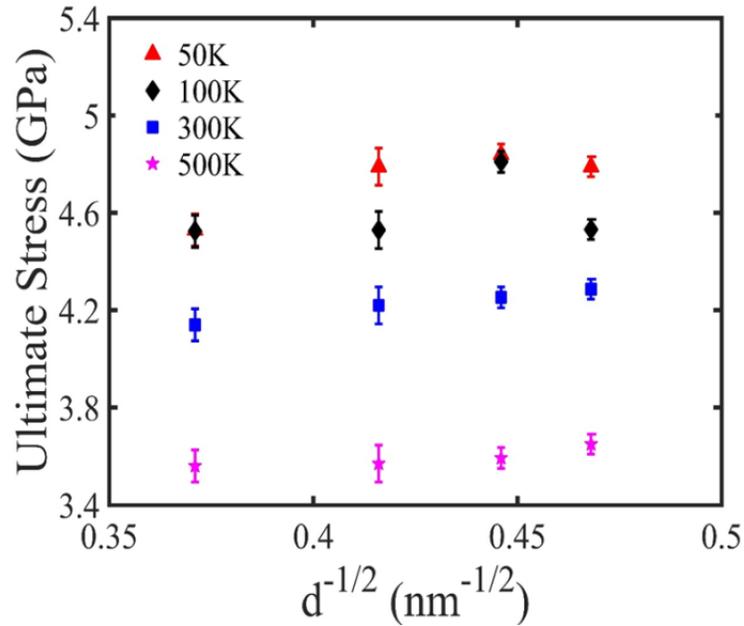

Fig. 3: Variation of ultimate stress with the change of average grain size of the *Al-Cu* nanostructure nanowire. The data presented here with the error bar is calculated from 10

different sample. At high temperature, ultimate stress shows increasing pattern with the decrease of grain size as a result of Hall-Petch effect.

We also investigated the flow stress of the nanostructure at different temperatures to ensure the Hall-Petch and inverse Hall-Petch finding at high and low temperature respectively. The flow stress for different grain sizes and temperatures are shown in Fig. 4. To calculate the average flow stress, the mean value of flow stresses from strain value of 8% to 12% for the nanostructure.

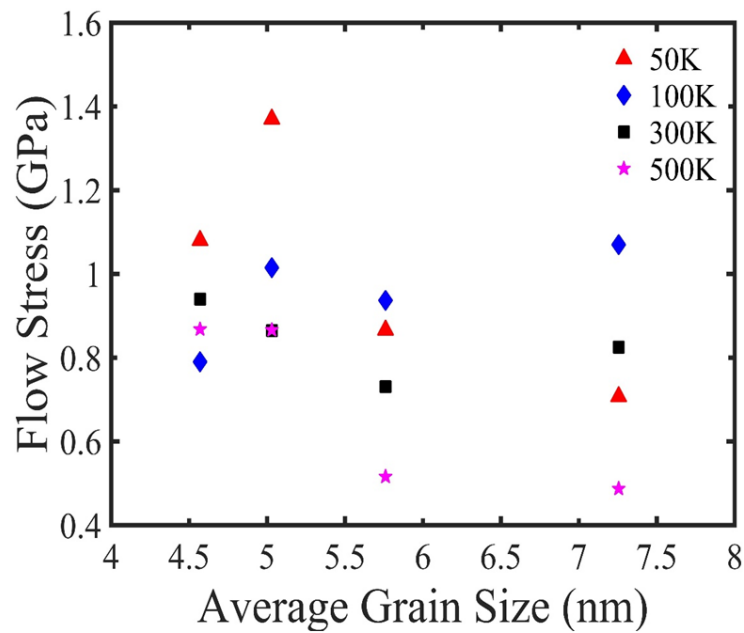

Fig. 4. Variation of average flow stress with the average grain size of the *Al-Cu* nanocrystalline nanostructure at different temperature. Here, the flow stresses shown are the average value of stresses from strain values 8 to 12%.

At the higher temperature of 300K and 500K, the flow stresses decrease with the average grain size of the material. To be coherent with Hall-Petch effect, this is favorable that the flow stress should be decreased with the average grain size[39]. At a low temperature of 50K and 100K, we see that the flow stress initially increase and show a higher value for the grain size of 5.03 nm and then decreases again. This finding corroborates that there is inverse Hall-Petch effect at low temperatures after a critical grain size.

## 3.2. Deformation mechanism

In nanocrystalline materials, the deformation is governed by the GB sliding, GB diffusion, dislocation nucleation and gliding, intragranular dislocation climb and elastic deformation in the grain[40]. At high temperatures of 300K and 500K, we observed the same deformation mechanism for different grain sizes in *Al-Cu* nanostructure nanowire. Among the grains of *Al* and *Cu*, an Aluminum grain is weaker and the failure occurs at that particular *Al* grain. This happens because *Al* grains are relatively weak when the shear force acts on it due to the normal tensile loading. The phenomenon of the weakening of a particular *Al* grain can be visualized in Fig. 5 by atomic configurations of *Al-Cu* nanostructure of average grain size 7.26 nm at 300K temperature. In Fig 5(b), the partial dislocation (shown by red atoms) formation is visible which is confined inside the weak grain. In Fig. 5(c), we can see that atoms diffuse from surrounding grains and that particular weaker grain become completely amorphous. The stress also accumulates along the GB of that particular grain. Now the important question is, which *Al* grain should initiate failure first? This depends on the orientation of the grain and loading condition. The failure may occur at different positions of different grains. From the weak grain, the formation of a void is initiated as shown in Fig. 5 and after that, the nanostructure behaves like ductile materials showing flow stress and necking behavior. This is the failure mechanism for the case of all grain size at high temperatures (300K-500K).

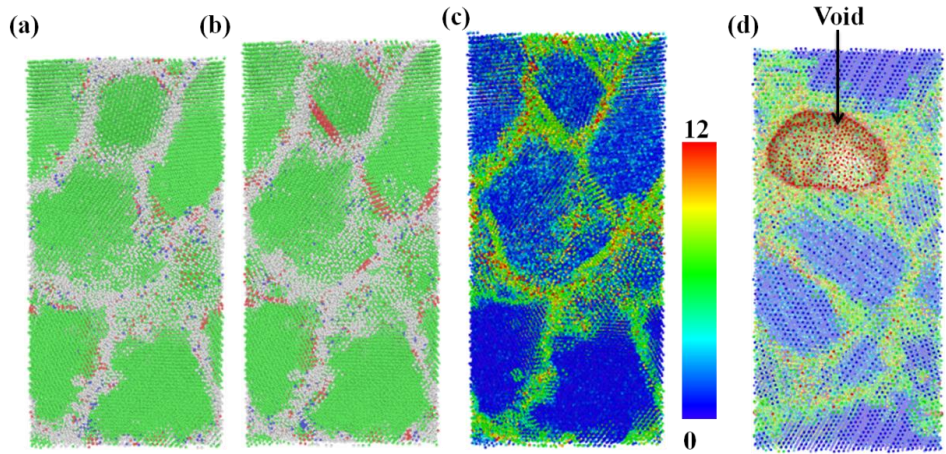

Fig. 5: Snapshots of atomic configuration for *Al-Cu* nanostructure of 7.26 nm grain size at (a) 0% strain, (b) 3.8% strain at a temperature of 300K. The green, red and white atoms shown in (a) and (b) denotes fcc, hcp and other atomic configuration. (c) centrosymmetry parameter at strain of 6.3%. The legend shows the centrosymmetry values.(d) void formation in a particular grain.

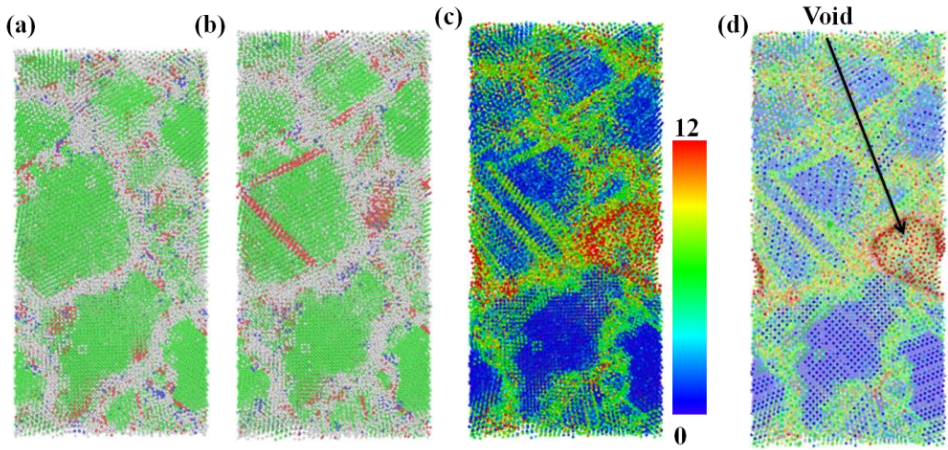

Fig. 6: Snapshots of sample configuration for *Al-Cu* nanostructure of 5.76 nm grain size at (a) 0% strain, (b) 5.5% strain at a temperature of 300K. The green, red and white atoms shown in (a) and (b) denotes fcc, hcp and other atomic configuration. (c) centrosymmetry parameter at strain of 7%. The legend shows the centrosymmetry values.(d) void formation in a particular grain.

The explanation of Hall-Petch behavior can be understood from the Fig. 6. For illustration, an average grain size of 5.76 nm is visualized with atomic configurations and can be compared with Fig. 5. The ultimate stress is higher for a grain size of 5.76 nm compared to 7.26 nm due to the formation of Lomer-Cortell lock (see Fig. 6(b)) marked by the red atoms. To break this lock, higher stress is required and this makes the ultimate stress higher for grain size of 5.76 nm compared to that of 7.26 nm grain size. Similar lock formation and dislocation activities are observed for smaller grain sizes (4.57 and 5.03 nm) at a temperature of 300K and 500K.

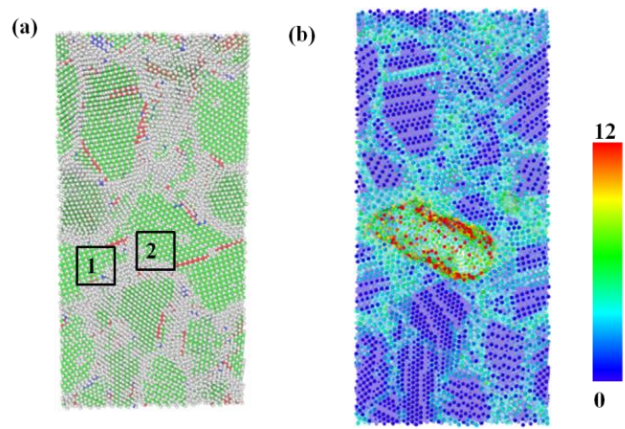

Fig7: Snapshots of sample configuration for *Al-Cu* nanostructure at 4.57 nm grain size at (a) 0% strain at a temperature of 100K. The green, red and white atoms shown in (a) denotes fcc, hcp and other atomic configuration. (b) Grain 1 and 2 fails at the same time and create a void inside the nanostructure. The color legend shows the centrosymmetry value for (b).

Then we found that down to a critical grain size of 5.03 nm, *Al-Cu* nanocrystalline nanostructure fails in a manner following Hall-Petch relation at lower temperatures (50K and 100K). After that, for a smaller grain size of 4.76 nm, the trends of ultimate stress with respect to grain size is inversed following inverse Hall-Petch relation. At low temperature, the materials show more brittle nature. That is why the grain boundary simultaneously affects two distinct grains of

Aluminum and two grains fail at a time. As a result, the ultimate stress decreases for low grain size at low temperature unusually (see Fig. 3). Here, both the grain size and temperature come into playing a role in the failure of the nanostructure because the volume of the grain boundary is higher as the grain size is low and at the same time, low temperature resists the sliding of the GB. As a result, two grains deform simultaneously (see Fig. 7) which reduces the ultimate stress of the nanostructure nanowire.

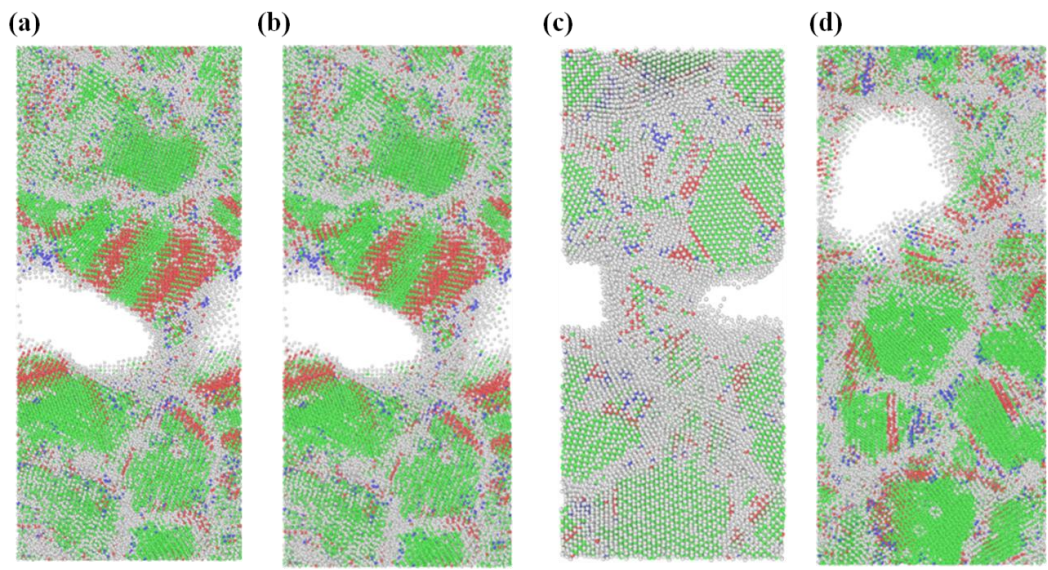

Fig 8: Failure of the *Al-Cu* nanostructure at 15% strain for (a) 7.26 nm (b) 5.76nm, (c) 5.03nm and (d) 4.57 nm at 300K temperature. For all the grain size the void formation is visible. The green, red and white atoms shown denotes fcc, hcp and other atomic configuration.

### 3.3. Effect of strain rate

To investigate the effect of strain rate on the tensile simulation we varied the strain rate from $10^{11}$ to $10^8$ s$^{-1}$ keeping the temperature fixed at 300K for a particular grain size of 7.26 nm. From Fig. 9, it is clear that at higher strain rate, the ultimate stress is higher. We fit the ultimate stress and strain rate taking its logarithmic value. The data fitted shows a pattern of a straight line with the equation shown in the Fig. 9. From the fitted equation, it is easy to construe the ultimate stress for any strain rate at 300K temperature.

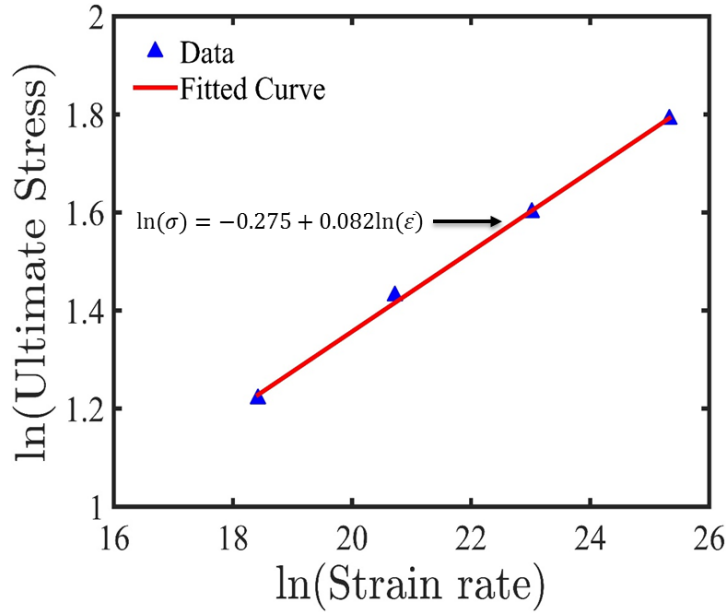

Fig 9: Variation of ultimate stress with the strain rate for grain size of 7.26 nm and temperature of 300K. The equation is derived by fitting the data which shows a positive slope.

### 3.4 Dislocation nucleation and propagation

In Fig. 10, the variation of dislocation density with respect to strain for different grain sizes is shown. It is visible that during the failure time, the dislocation density reduces significantly. It can be explained by the grain behavior at the failure strain. At failure strain of around 0.64-0.67, a grain becomes completely amorphous reducing the number of dislocation line. At the initial stage, the dislocation line is higher but at the failure strain, the dislocation line reduces as the grain is changed to amorphous nature. We observe that the type of dislocation is mainly Shockley partial type (1/6<112>). Two of this type of Shockley partial dislocation can be combined into a perfect dislocation (1/6<112> +1/6<112> =1/2<110>). As the slip plane in FCC crystals is {1 1 1}, there should be the formation of perfect and partial dislocation which is manifested in our results.

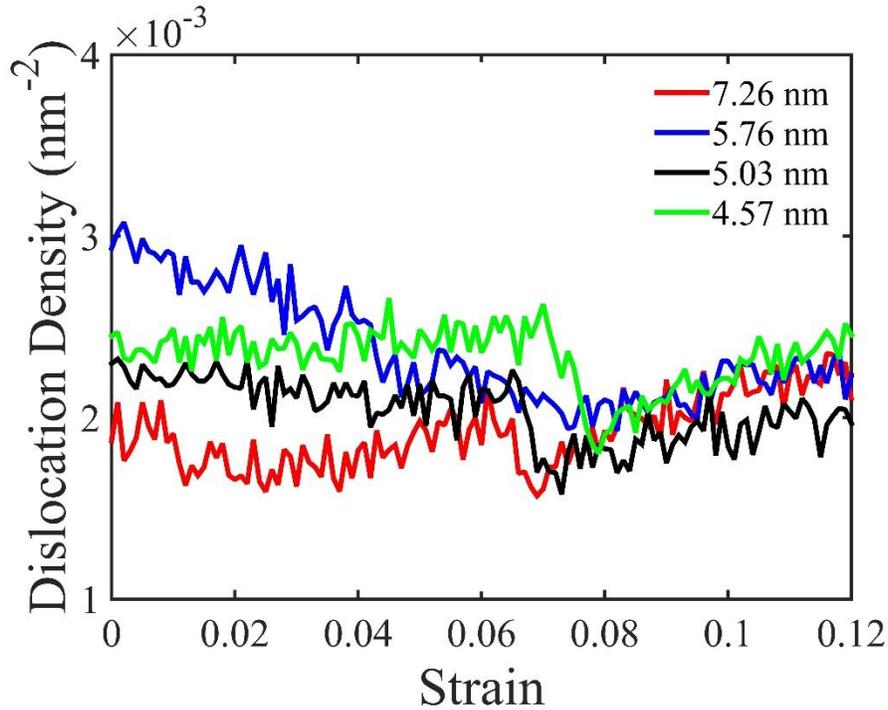

Fig. 10: Variation of dislocation density at different strain for various grain size. Dislocation density decreases at the fracture region significantly.

At smaller grain size, the dislocation line length is shorter. From Fig. 11(a), the surface defect mesh can be seen which is constructed from the GB. Dislocation line becomes discontinuous due to this GB. As the material is deformed (See Fig. 11(b)), the grain size is increased. The grain tries to accumulate the dislocation and as a result dislocation length increases and sometimes their interaction with some other dislocation inside the same grain can annihilate the dislocation. At the failure point, a particular *Al* grain becomes amorphous which causes the reduction of total dislocation density on the material.

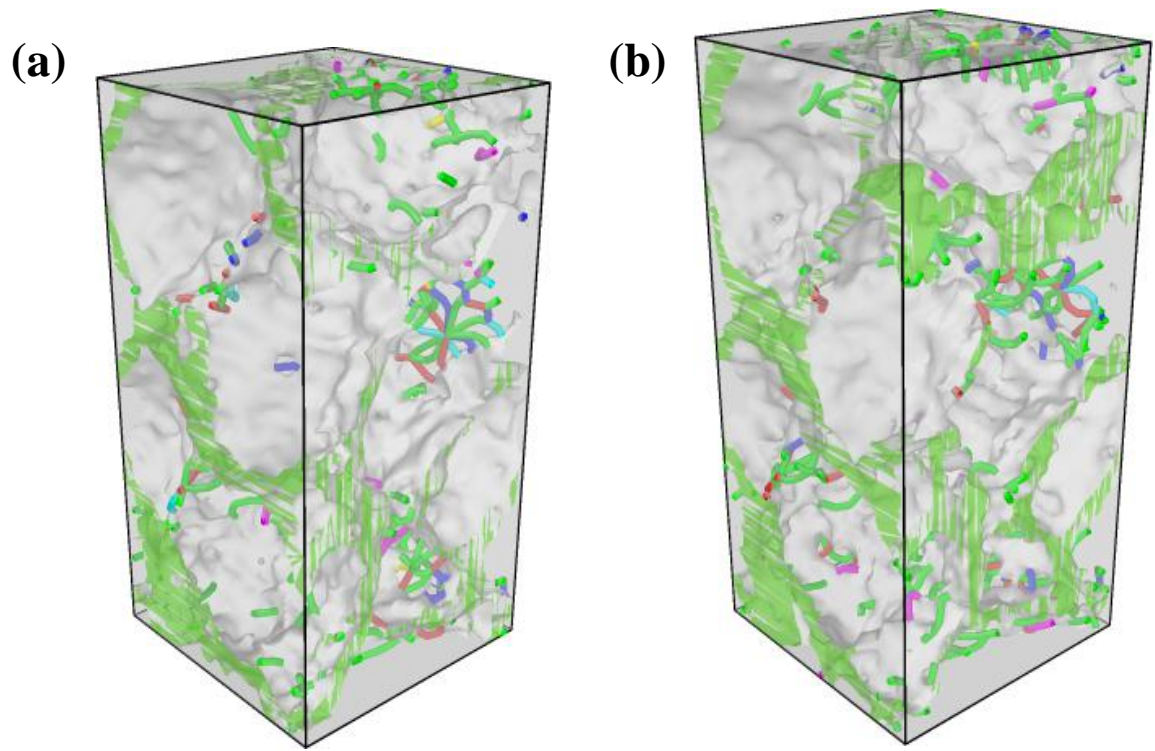

Fig. 11: Dislocation line with surface defect mesh (a) at the initial stage (b) during failure for grain size of 7.26 nm and 300K temperature.

## 4. Conclusions

The grain size and the temperature effect on the deformation mechanism of nanocrystalline *Al-Cu* nanostructure are investigated in this paper. The following key findings can be drawn from the results discussed above:

- The *Al-Cu* nanostructure manifests Hall-Petch effect in the smaller grain size of 4.57 to 7.26 nm at a higher temperature of 300K to 500K. The deformation mechanism shows a particular Al grain becomes weaker and fails upon the application of load. The Hall-Petch behavior is also supported by the flow stress.

- At higher temperatures (300-500K) and lower grain sizes (4.57 -5.76 nm), there is formation of Lomer-Cortell lock. This increases the ultimate stress of the nanostructure.

- At temperatures of 50 to 100K, two grains fail simultaneously and the ultimate stress decreases for a grain size of 4.57 nm. This gives rise to the evidence of inverse Hall-Petch effect at low temperature.

- The deformation is governed by the dislocation nucleation and its arrest in a grain, sliding of GB, and elastic field of the dislocation cores. The dislocation activity is reduced at the failure of the material.

These results can be further useful for the investigation of *Al-Cu* nanostructure at the nanoscale.

**Acknowledgment:**

The authors gratefully acknowledge the computational facilities provided by the Department of Mechanical Engineering, Bangladesh University of Engineering and Technology, Dhaka-1000.